\documentclass[12pt]{spieman}  
\usepackage{amsmath,amsfonts,amssymb}
\usepackage{graphicx}
\usepackage{setspace}
\usepackage{tocloft}
\usepackage{lineno}
\usepackage{multirow}
\usepackage[table,xcdraw]{xcolor}

\title{Reconstructing galactic feedback history via the Lyman-$\alpha$ forest  with Habitable Worlds Observatory}

\author[a,*]{Megan T. Tillman}
\author[b]{Joseph N. Burchett}
\author[a,c]{Blakesley Burkhart}
\author[d]{Vikram Khaire}
\author[e]{Sanchayeeta Borthakur}
\affil[a]{Department of Physics and Astronomy, Rutgers University,  136 Frelinghuysen Rd, Piscataway, NJ 08854, USA}
\affil[b]{New Mexico State University, Department of Astronomy, Las Cruces, NM, 88011}
\affil[c]{Center for Computational Astrophysics, Flatiron Institute, 162 Fifth Avenue, New York, NY 10010, USA}
\affil[d]{Department of Physics, Indian Institute of Technology Tirupati, Tirupati, Andhra Pradesh, 517619, India}
\affil[e]{School of Earth \& Space Exploration, 
Arizona State University, 781 Terrace Mall, Tempe, AZ 85287, USA}

\newcommand{\lya}{Ly$\alpha$}

\cftpagenumbersoff{figure}
\cftpagenumbersoff{table} 
\begin{document} 
\maketitle

\begin{abstract}
Recent studies have focused on the low-$z$ Ly$\alpha$ forest as a potential constraint on galactic feedback, as different AGN and stellar feedback models in hydrodynamic simulations produce varying intergalactic medium (IGM) statistics.
However, existing low-$z$ Ly$\alpha$ forest data provide insufficient observational constraints for simulations due to their low precision and the lack of observations from $z \sim 0.4$ to $1.8$.  
The Habitable Worlds Observatory, equipped with an ultraviolet (FUV / NUV) spectrograph, could provide transformative data for this science by increasing both absorbers in the redshift range where current data are lacking and increasing the precision of the Ly$\alpha$ forest observational data at $z \leq 0.4$.
This spectrograph should cover the wavelengths $1215-3402$ \AA~and have a spectral resolution of $R =$~40,000.
Distinguishing between certain active-galactic nuclei feedback models requires high precision ($\sim 5-10$\%) measurements of the Ly$\alpha$ forest 1D transmitted flux power spectrum for $z\leq 1.8$.
This requires $\sim 830$ QSO spectra with S/N~$\geq5$ - $25$ depending on redshift. 
This data would also enable a comprehensive study on the thermal state of the IGM across time, and address the tension between the observed and simulated Ly$\alpha$ forest $b$-value distribution. 
\end{abstract}


{\noindent \footnotesize\textbf{*}Megan Tillman,  \linkable{mtt74@rutgers.edu} }

\begin{spacing}{2}   

\section{Introduction}
\label{sect:intro}

Shortly after the discovery of the first quasistellar object or quasar (QSO) more than half a century ago, studies showed a trough in the observed spectrum blueward of the Lyman-$\alpha$ (\lya) emission line \cite{Gunn+Peterson1965, Bahcall+Salpeter1965}. 
This trough was the result of the absorption of photons by neutral hydrogen (HI) along the line of sight between the QSO and Earth. Several years passed before the realization that the \lya~absorption feature was, in fact, not a single saturated region but a series of individual absorption lines comprising what we now refer to as the \lya~forest \cite{Bahcall:1971}. 
Each individual absorption line in the forest represents an absorbing gas complex in the intergalactic medium (IGM) with observed wavelength $\lambda_{\rm obs} = 1215.67 (1 + z)$\AA~tracing neutral hydrogen\cite{Oort:1983}.

Studying how \lya~forest statistics evolve from high ($z > 2$) to low ($z \sim 0.1 $) redshift can reveal what factors set the thermal state of that gas and thus the IGM, after reionization \cite{McQuinn:2016}. 
Multiple mechanisms can affect the forest, from assumed cosmology to the predicted ionizing radiation background; even galactic feedback can affect the forest \cite{Dave:1999, Kollmeier:2014, Christiansen:2020}.
At high redshift ($3<z<6$), the structure of the \lya~forest is assumed to be dominated by cosmological factors, such as the nature of dark matter\cite{Palanque-Delabrouille:2020,Villasenor:2023}, with temperature and neutral fractions set by the uniform ultraviolet background (UVB)\cite{Khaire:2019UVB, Puchwein:2019UVB, Faucher-Giguere:2020UVB}. At low redshift, feedback is expected to play a greater role with more observed AGN activity \cite{BroderickI:2012, ChangII:2012}. 
Determining to what extent each of these mechanisms plays a role is difficult due to their degeneracies, as many AGN feedback effects may be accounted for in an ionizing background model \cite{Mallik:2023}. 
Furthermore, additional heating in the IGM may be accounted for in other non-traditional heating mechanisms such as dark photons which might show different redshift evolutions than AGN feedback effects \cite{Bolton:2022}.
Several attempts have been made to unravel the true impact of feedback on the low-$z$ \lya~forest in recent years \cite{Viel:2017, Gurvich2017, Tonnesen:2017, Nasir:2017, Christiansen:2020, Chabanier:2020, Burkhart_2022, Khaire:2022, Dong:2023, Khaire:2023, Dong:2024, Khaire:2024}, however, the true role of feedback remains under question.
Studying the evolution of \lya~forest statistics across a wide range of redshifts has great potential to reveal the role that galactic feedback may play in influencing the thermal state of the IGM and the effort to disentangle the degeneracies with the ionizing background.

The ultraviolet (UV) spectrographs aboard the Hubble Space Telescope (HST), namely the Cosmic Origins Spectrograph (COS) and Space Telescope Imaging Spectrograph (STIS), have obtained nearly 900 QSO spectra covering a wavelength range of $\sim 1150-3200$ \AA. 
This wavelength range, in principle, covers the \lya~forest from $z=0$ to $z\sim1.6$. To build large samples for \lya~forest statistical analysis, a large catalog of QSOs is required across a wide redshift range. 
From observations, there are about 80 QSOs probing the \lya~forest at $z < 0.5$ \cite{Danforth:2016, Khaire:2019 ,Khaire:2023}. For $0.5 < z < 1.7$, there are $\sim45$ QSOs currently public in the HST archive and another 15 to be public by 2026 probing the \lya~forest observed with STIS (E230M) and COS gratings G185M and G225M. 
Given the currently available data, conducting science using the \lya~forest at $z < 1.8$ is limited to $\sim20-25$\% precision on the 1D transmitted flux power spectrum for $z\leq0.41$\cite{Khaire:2019}, a statistic often used since it encodes information on the thermal state of the IGM. 
Collecting additional data, especially in the redshift range $0.41 < z < 1.8$, will be essential for further study of the \lya~forest, and thus the IGM, during a time when feedback effects may become dominant. 
However, the number of remaining known QSOs that are bright enough in the UV to be feasibly observed with COS and STIS is too few to make substantial improvements on existing statistical constraints.

Once sufficient empirical statistics are obtained, we may compare these observations with cosmological hydrodynamical simulations that implement different cosmologies, UVB, and feedback models to help distinguish between these models and reveal what mechanisms set the thermal state of the low-$z$ IGM.
These cosmological simulations are unable to simultaneously simulate small-scale physics driving feedback, the ejection of matter and energy from active galactic nuclei (AGN), stellar winds, and supernovae, and larger-scale evolution of galaxies and the cosmic web.  
Therefore, feedback effects are modeled with `subgrid' prescriptions that are in turn tuned to reproduce galaxies resembling those observed; the resulting physical conditions of the IGM follow as predictions of the simulations. 
Since many feedback models can reproduce similar galaxy statistics such as stellar or halo mass functions, analyzing the predicted IGM through the \lya~forest may provide a novel method for constraining these different models.

Given the limited capacity to grow observational datasets with HST, taking a transformational step in \lya~forest statistical analysis will likely require a future UV mission.  
The Habitable Worlds Observatory (HWO) flagship mission could enable such a transformational advance in studying the \lya~forest at $z<2$ if equipped with a high-resolution UV spectrograph. 
In this work, we evaluate the requisite instrument specifications and observing strategy for distinguishing between feedback models of a state-of-the-art cosmological simulation.

\section{Methods}

    Herein, we take a forward modeling approach where we employ the Simba \cite{SIMBA} hydrodynamical cosmological simulation and produce synthetic observations of `sightlines' through the simulation volume to emulate observed QSO spectra in the real Universe.  
    We then directly measure observable quantities, such as the \lya~optical depth, and their resulting statistics, such as the 1D transmitted flux power spectrum (P1D), from these synthetic spectra.  
    This approach enables us to directly control parameters such as spectral resolution and sample sizes of QSOs that determine the resulting datasets' constraining potential.

    The Simba simulation \cite{SIMBA} is run using the GIZMO cosmological gravity plus hydrodynamics solver \cite{Hopkins:2015, Hopkins:2017} based on GADGET-3 \cite{Springel:2005}.
    Radiative cooling and photoionization heating are modeled using GRACKLE-3.1 \cite{Grackle:2017}, which includes metal cooling and non-equilibrium evolution of primordial elements. 
    The simulation used herein has a comoving box length of 100 Mpc/h with $1024^3$ dark matter particles and $1024^3$ gas elements.
    Simba implements stellar and AGN feedback with an AGN jet mode that has a substantial impact on IGM temperature \cite{Christiansen:2020, Tillman:2023AJL, Tillman:2025}. 
    The AGN jet feedback specifically has large-scale impacts that affect the abundance of neutral hydrogen in the \lya~forest \cite{Tillman:2023AJL,Tillman:2023AJ,Tillman:2025}.
    
    Synthetic spectra are generated from the Simba simulation\cite{SIMBA} using the publicly available code Fake Spectra Flux Extractor (FSFE) \cite{Bird:2015, Bird:2017, Qezlou:2022}.
    The FSFE package is fast, parallel, and is written in C++ and Python3 with a Python-based user interface. 
    For the work presented here, 5,000 sightlines have been found to be sufficient for the convergence of various \lya~statistics, including the P1D. 
    Given the finite box size of the simulation, having enough sightlines is essential to obtain a representative population of \lya~absorbers. 
    The sightlines are randomly placed in the simulation box along the z-axis. 
    The optical depths along these sightlines are output by FSFE from which the flux for the analyses herein is calculated as $F(v) = e^{-\tau}$, where $v$ is the velocity along the line of sight in units of km/s.
    
    For analyzing variation with redshift pathlength, we generate spectra as described above but with a different number of sightlines. 
    This allows for an analysis on the number of QSO observations necessary to obtain converged data for the \lya~forest.
    We estimate the error from varying the total path length with a bootstrapping technique.
    We resample the original 5000 spectra $N=5000$ times and take the 95\% confidence interval to predict the error within 2$\sigma$.
    We discuss the results to achieve 1$\sigma$ confidence; however, we present the 2$\sigma$ results, as we expect the estimated error to be under-predicted due to cosmic variance and the limited box size of the simulation.

    To analyze the effect of the spectral resolution of the telescope, we generated the same 5,000 sightlines but convolved the spectra with a Gaussian line spread function (LSF). 
    Although the HWO spectrograph may have an LSF that deviates from Gaussian, as does that of COS\cite{HST-COS}, instrument designs typically strive for a well-behaved Gaussian LSF.
    To analyze the effects of signal-to-noise ratio (S/N), a Gaussian noise vector is calculated from the flux variance and desired S/N. 
    The P1D is calculated separately for the generated sightlines and the noise vector. 
Then the sum of the two represents the \lya~forest P1D that varies with S/N.
    The FSFE package includes the infrastructure necessary to produce all of these spectral variations, either by regenerating the optical depths over the given pathlengths and spectral resolutions or by adding noise in post-processing.

    From the spectra, we analyze the 1D transmitted flux power spectrum. 
    The P1D is computed as the 1D Fourier transform of the flux fluctuations $\delta_F(v)$, with

        \begin{equation}
            \delta_F(v) \equiv \frac{F(v)-\langle F\rangle}{\langle F\rangle}.
        \end{equation}

    \noindent The mean transmitted flux $\langle F \rangle$ is averaged over all sightlines. 
    The corresponding P1D is then expressed in terms of the wavenumber $k$ in units of s/km. 
    For the P1D, optical depths greater than $\tau = 10^6$ are filtered out of the spectra because they are considered damped \lya~regions (DLAs) and render large regions of the observable pathlength unusable for \lya~forest analysis.
    This avoids having to deal with the damped wing regions of the spectra associated with DLAs.
    The entire spectral line is thrown out of the analysis if it contains optical depths hitting the threshold. 
    Less than 1\% of the artificial spectral lines are thrown out in this way, so this filter will not have significant impact on the overall results.
    In fact, the impact on the resulting P1D is less than a few percent.
    Studies on DLA regions in Simba at higher redshifts ($z=3$ to 5), find that Simba underproduces DLAs compared to observations \cite{Hassan:2020}.
    This likely propagates to lower redshifts if the result is due to the AGN feedback model, which is more impactful at lower $z$.
    Conversely, the under-prediction could be due to cosmic variance and the limited simulation box size.
    Regardless, the number of QSO observations required based on the Simba simulation is likely underestimated.

\section{Results}

    The \lya~forest P1D statistic encapsulates diverse environments. The \lya~forest traces both small and large scale structure of the cosmos. As such, the P1D has been used to form cosmological constraints such as the nature of dark matter and the mass of neutrinos \cite{Yeche_2017,Palanque-Delabrouille:2020, Villasenor:2023}.
    On small physical scales ($k \gtrsim 0.04$ s/km), the P1D reveals information on the thermal state of the IGM, allowing for constraints on the ionizing background and reionization \cite{Boera:2019, Zhu:2021}.
    Studies have shown through simulations that the effect of AGN feedback on the predicted P1D, as seen through simulations, has the potential to introduce bias in the measured cosmology \cite{Viel:2013, Chabanier:2020, Burkhart_2022}.
    Furthermore, the impact of AGN feedback models on the predicted P1D from various simulations has shown the greatest potential to isolate the impact of galactic feedback \cite{Burkhart_2022, Khaire:2024, Tillman:2025}.

    \begin{figure}
        \begin{center}
        \begin{tabular}{c}
        \includegraphics[height=9.5cm]{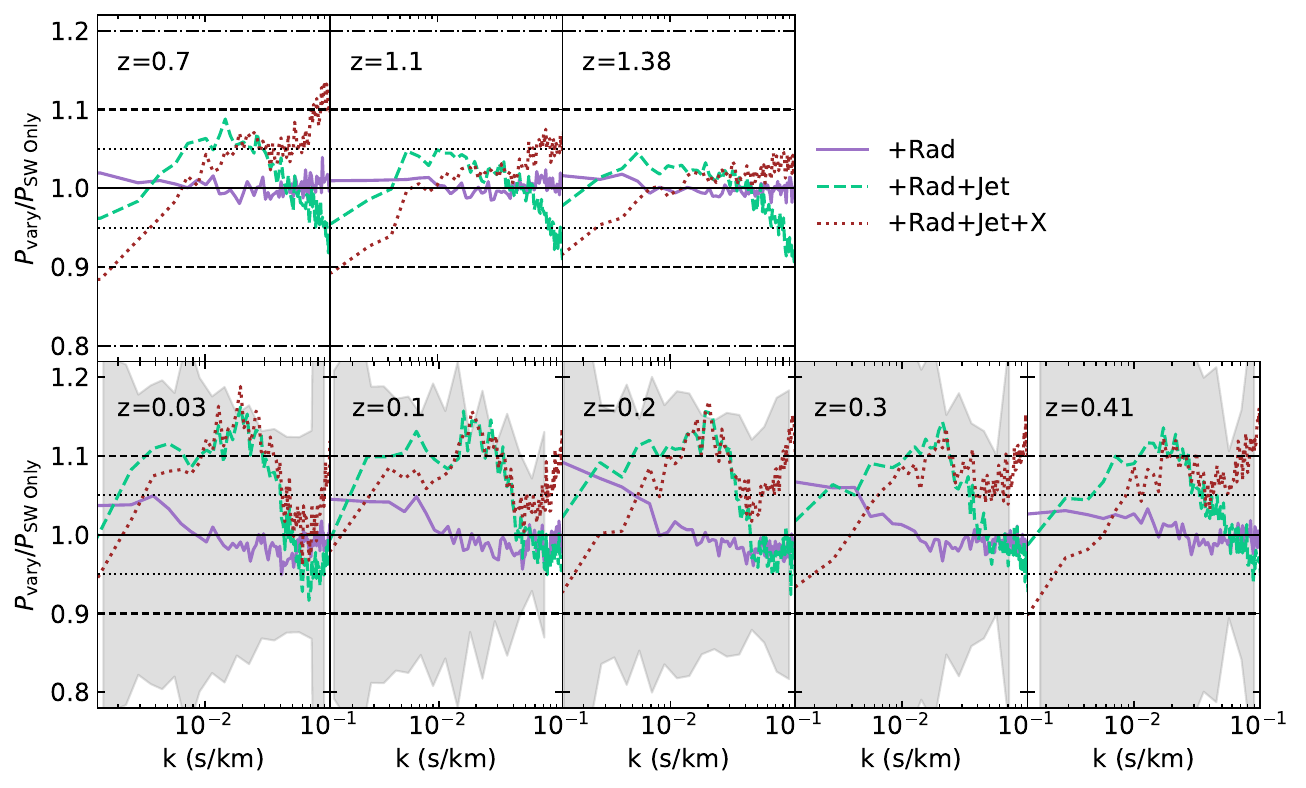}
        \end{tabular}
        \end{center}
        \caption 
        {\label{fig:P1Dadapted}
        Figure adapted from recent work comparing different AGN feedback implementations of the Simba simulation model \cite{Tillman:2025}. Shown is the P1D ratio of different AGN feedback implementations (+Rad, +Rad+Jet, and +Rad+Jet+X) to the stellar feedback (SW) only case. Each P1D is rescaled to have the same mean transmitted flux as the SW only case which is done to account for uncertainties due to the unconstrained nature of the ionizing background. The shaded regions for $z\leq0.41$ represent current observational error bars \cite{Khaire:2019}. Dashed and dotted lines mark 10\% and 5\% P1D precision respectively. For $z\leq0.41$, 10\% precision can distinguish between the SW only case and the +Rad+Jet or +Rad+Jet+X case. For $z=0.7$, 10\% precision can distinguish between SW only and +Rad+Jet+X.}
    \end{figure}

    Recent work comparing Simba simulations with different implementations of AGN feedback found that observational data on the \lya~forest P1D with a precision of 10\% would be enough to distinguish between certain models \cite{Tillman:2025}.
    Fig.~\ref{fig:P1Dadapted} is an adapted figure from that work showing the AGN feedback variations and the predicted P1D for relevant redshifts.
    The solid black lines represent a simulation with supernovae feedback only (SW only) while the colored lines add in the different AGN feedback modes one at a time.
    The purple lines add in radiative feedback (+Rad), the teal lines add in the AGN jets (+Rad+Jet), and the maroon lines add in the X-ray mode (+Rad+Jet+X). 
    More information on the feedback modes and the simulation variants can be found in the Simba simulation presentation paper\ \cite{SIMBA}.

    Although 10\% precision is enough to distinguish between certain feedback variations at $z\leq 0.7$, 5\% precision would allow even further constraints on the feedback variations shown.
    The Simba simulations are relatively unique in their ability to study the impact of feedback on the \lya~forest in this way as the simulations studied only vary the feedback.
    Comparisons between Illustris and IllustrisTNG simulations, which implement different AGN feedback models, similarly show different impacts that AGN feedback has on the IGM \cite{Burkhart_2022, Khaire:2024}.
    There are also the HorizonAGN simulations that include a simulation run that removes AGN feedback to better study the effects of that feedback on galaxy formation and evolution \cite{HorizonAGN}. 
    A study examining $z\geq 2.0$ comparing these simulations found that the AGN feedback impacts the \lya~forest \cite{Chabanier:2020}. 
    
    The ability to study the \lya~forest in simulations is often limited by low resolution and small box size. 
    Works making such analyses make attempts to address these limitations, but bigger simulations with higher resolution, which still implement complex astrophysics such as feedback, are often too expensive to run. 
    This makes estimates on uncertainties approximate, especially when considering AGN feedback impacts.
    For example, the reduced box size of the Simba feedback variant simulations (50 Mpc/h side length as opposed to 100 Mpc/h) leads to variation in the halo mass function up to 15\% at $z=0$. 
    The full impact this will have on the results due to AGN feedback is unclear, with a rough estimate including a $\pm 9$\% variation for the 60\% impact seen in the P1D due to AGN jets at $z=0.1$ \cite{Tillman:2025}. 
    
    To account for uncertainties due to cosmic variance, it is helpful to aim for a precision higher than 10\% on the observational data collected.
    Pushing the \lya~forest P1D precision to 5\% would clearly distinguish between feedback models at $z\leq0.41$ and would allow a similar analysis for $0.7<z<1.38$, as shown in Fig.~\ref{fig:P1Dadapted}. 
    This would also provide room for additional uncertainties in the AGN feedback models and simulations.
    To enable such a measurement, in addition to more observations of the \lya~forest in the redshift range $0<z<1.8$, the spectrograph used will need to have sufficient spectral resolution and wavelength coverage. 

    \begin{figure}
    \begin{center}
    \begin{tabular}{c}
    \includegraphics[height=7cm]{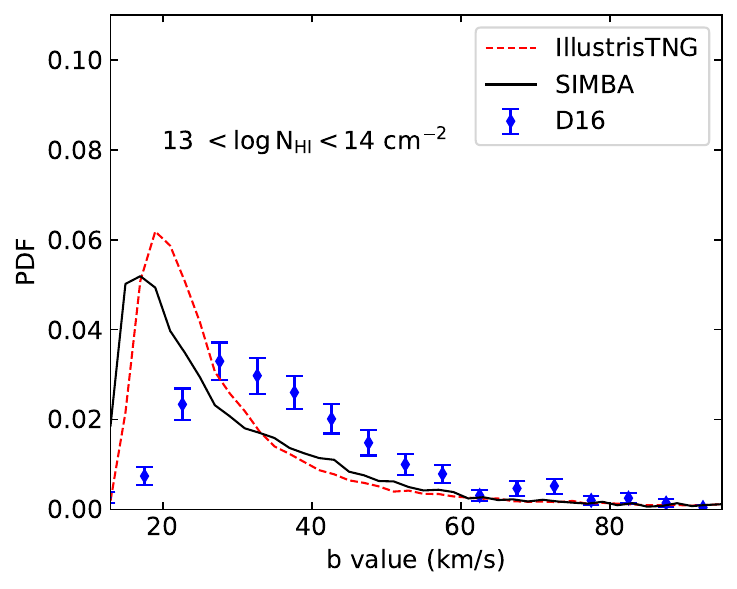}
    \end{tabular}
    \end{center}
    \caption 
    { \label{fig:bvalue}
    The predicted $b$-value from the Simba and IllustrisTNG simulations \cite{SIMBA, IllustrisTNG} versus the observed $b$-value distribution \cite{Danforth:2016}. The distribution is for absorbers with column densities of $10^{13}$ to $10^{14}$ cm$^{-2}$ as this is where observations are available. The simulations predict a higher number of low $b$-values ($b<30$ km/s) than what observations show.
    }
    \end{figure} 

    To observe the \lya~forest in the redshift range desired, the spectrograph must be able to observe wavelengths from 1215-3402\AA. 
    This would cover the \lya~transition up to $z=1.8$.
    While at higher redshifts, the \lya~transition can be observed from ground-based telescopes, the impacts of AGN feedback are not as prominent.
    Several studies have shown that the effects of galactic feedback on larger scales manifest themselves and become increasingly pronounced at $z<2$ \cite{Tillman:2023AJ, Saha:2024, Tillman:2025}. 
    
    In terms of spectral resolution, the minimum absorber width must be considered. 
    For the \lya~forest, absorbers can theoretically have $b$-values as small as 10 km/s corresponding to the average velocity of the hydrogen atom at $10^4 K$.
    To trust an absorption feature, resolving it with two resolution elements is ideal, which will require a high spectral resolution. 
    Many simulations predict an average $b$-value of $\sim 20$ km/s for the low-$z$ \lya~forest\cite{Viel:2017, Gaikwad:2017viper, Bolton:2021, Burkhart_2022,Tillman:2023AJ}, which is inconsistent with the current available data, which place the average value of $b$ closer to $\sim 30-35$ km/s \cite{Danforth:2016}.
    Fig.~\ref{fig:bvalue} shows a comparison of two simulations to the available observations to demonstrate this point.
    Proposed solutions to this discrepancy include nontraditional heating sources that have not been taken into account, such as galactic feedback or extra line-of-sight turbulent velocity components \cite{Viel:2017, Gaikwad:2017viper, Bolton:2021, Burkhart_2022, Tillman:2023AJ}.
    
    The spectral resolution of HST-COS is approximately 17 km/s, which hinders the resolution of absorbers with $b$-values below $\sim 35$ km/s.  Thus, it is possible that the true $b$-value distribution peaks at narrower lines.
    To resolve the $b$-value discrepancy and discern the salient heating mechanisms of the IGM, a higher spectral resolution instrument is paramount. 
    Insufficient spectral resolution will severely limit the utility of the next-generation space-based \lya~forest observations for analyzing small-scale structure and physical mechanisms. 

    \begin{figure}
    \begin{center}
    \begin{tabular}{c}
    \includegraphics[height=7cm]{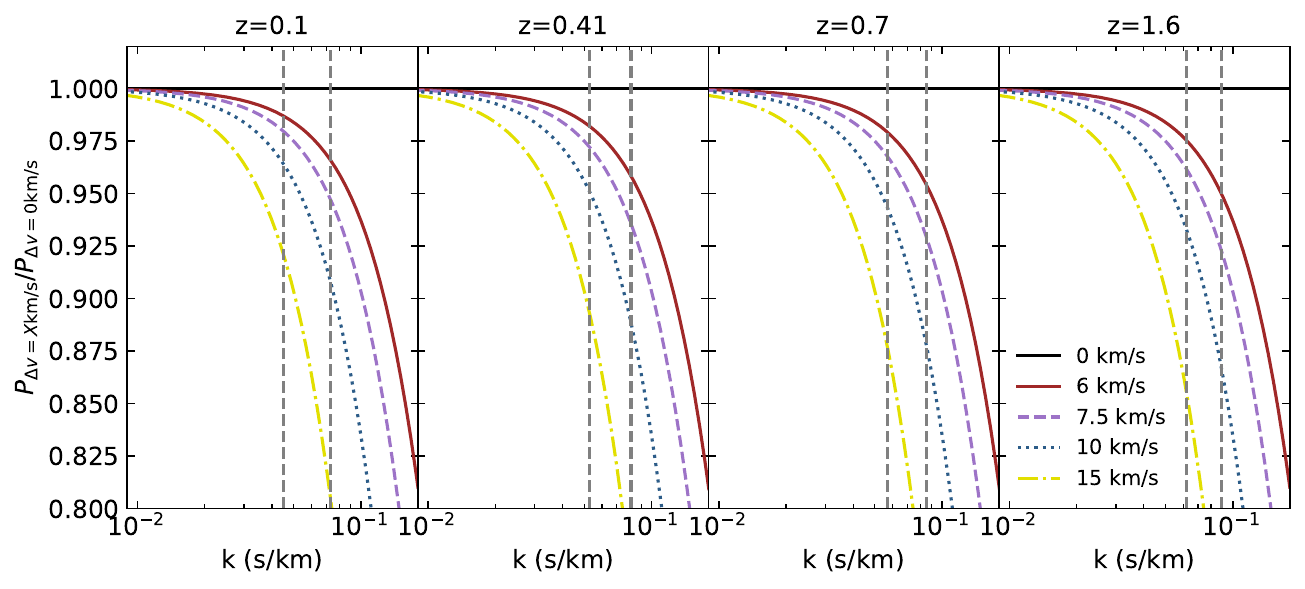}
    \end{tabular}
    \end{center}
    \caption 
    { \label{fig:P1Dspecres}
    The variation of the predicted 1D transmitted flux power spectrum (P1D) given different spectral resolutions within 4 redshift bins. On the y-axis is the ratio of the predicted P1D varying the instrument's spectral resolution (colored lines) to the predicted P1D assuming infinite resolution (black line). The x-axis is the $k$ value corresponding to the line of sight velocity. The dashed vertical lines represent the Jeans scale and a conservative estimate for the filtering scale accounting for pressure smoothing and thermal broadening, bracketing the relevant $b$-value scale.
    To constrain the P1D within a desired precision at some redshift, HWO's UV spectral resolution should be such that the corresponding curve lies within the precision between the dashed lines. For example, at $z=0.7$, 6 km/s resolution is required to constrain the P1D to within 5\% at the relevant scales. } 
    \end{figure} 
    
\begin{figure}
        \begin{center}
        \begin{tabular}{c}
        \includegraphics[height=7cm]{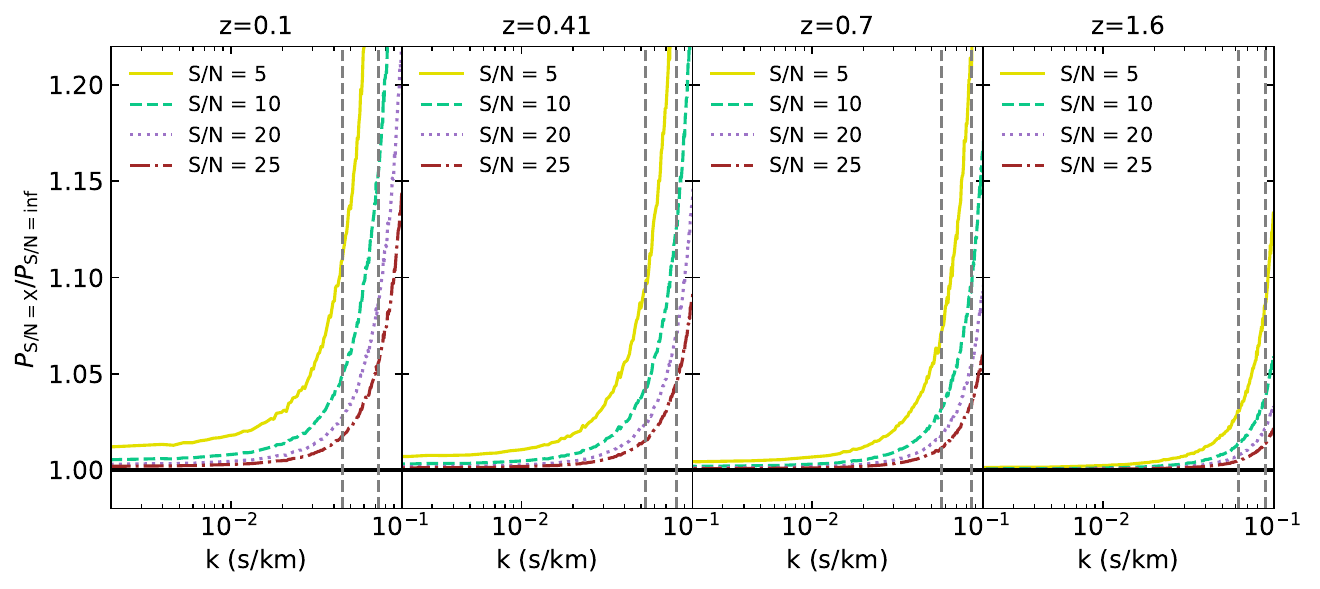}
        \end{tabular}
        \end{center}
        \caption 
        {\label{fig:P1Dsnr}
        Same as Fig. \ref{fig:P1Dspecres} but for variation in the S/N per resolution element of the spectral lines. The calculations were done assuming the instrument would have a spectral resolution of R=40,000. To constrain the P1D within the desired precision at some redshift, the S/N of the obtained observations should be such that the corresponding curve lies within that precision. The dashed vertical lines mark the relevant scale for analyses on the $b$-value.}
    \end{figure}

\begin{figure}
        \begin{center}
        \begin{tabular}{c}
        \includegraphics[height=7cm]{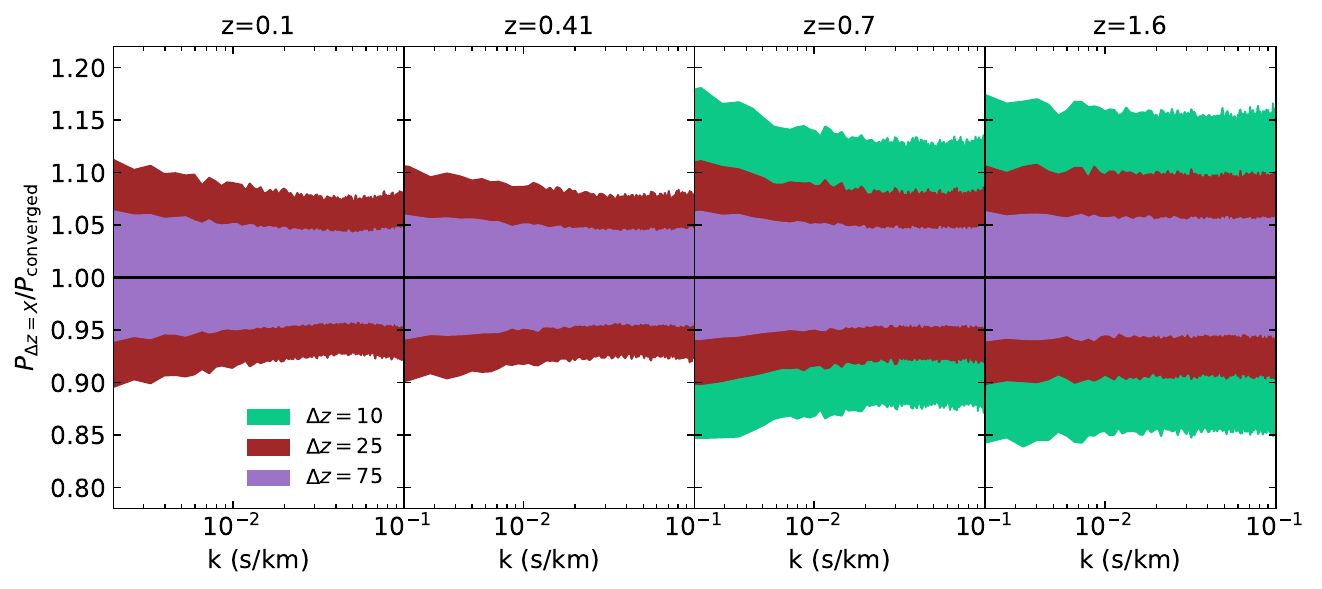}
        \end{tabular}
        \end{center}
        \caption 
        { \label{fig:P1Dpathlength}
        Same as Fig.~\ref{fig:P1Dspecres} but for variations in the predicted P1D given different total redshift pathlengths covered. The black line corresponds to the 5000 sightlines, which we consider to be converged. To constrain the P1D within the desired precision at some redshift, the redshift pathlength of the obtained observations should be such that the corresponding shaded region lies within that precision. The error is estimated through a bootstrapping technique and the shaded region represents a 95\% ($2\sigma$) confidence interval. The number of QSO observations required depends on $\Delta z$ and the width of the redshift bin in question. These uncertainties are likely underestimated since they were calculated in a finite simulation box size with on realization of the density field.}
    \end{figure}

    Fig.~\ref{fig:P1Dspecres},~\ref{fig:P1Dsnr}, and~\ref{fig:P1Dpathlength} all show ratios of the P1D varying different aspects of the hypothetical observations.
    The plotted ratios are the predicted P1D given a certain variation in observation parameters to the P1D given perfect conditions (i.e.\ infinite spectral resolution, S/N, etc.).
    These plots explore the impact of spectral resolution ($\Delta v$), S/N, and redshift pathlength ($\Delta z$) on the \lya~forest P1D to better inform what is required of both the spectrograph and the data it collects.
    
    Fig.~\ref{fig:P1Dspecres} shows the effect of different spectral resolutions on the predicted P1D.
    Spectral resolution affects the high $k$ end of the P1D which corresponds to small-scale structures.
    This end of the P1D is set by the temperature of the IGM and the $b$-value distribution of \lya~forest absorbers. 
    To make a definitive statement about where the peak of the observed $b$-value distribution is, 5\% precision is required up to the filtering scale. 
    
    The filtering scale, also known as the smoothing scale, is important to determine the relative number of high $b$-value to low $b$-value absorbers \cite{Hui+Rutledge:1999}.
    On this scale, linear gas fluctuations are smoothed because of finite gas pressure.
    For the IGM, this scale is not the Jeans scale when taking into account reionization history.
    The filtering scale is affected by the entire IGM thermal history, making calculating an exact value difficult.
    After reionization, the filtering scale is smaller than the Jeans scale and can be calculated from the Jeans scale using the factor $f_J$, which is between 0 and 1 and is generally smaller closer to reionization \cite{Gnedin+Hui:1998}.
    
    For our purposes, we make a conservative estimate for the filtering scale using $f_J=0.5$ and assume a gas temperature of $10^4$ K which allows us to determine up to what $k$ value we need to collect data for (mimicking a similar analysis done in previous work\cite{Tillman:2025}). 
    This is because for $z\leq 1.6$, the value for $f_J$ is likely to be larger than 0.5, pushing the $k$ requirement to larger scales that are less demanding in their spectral resolution requirements.
    Also included in the estimate for the filtering scale is a component that accounts for thermal broadening. 
    These components are added in quadrature to make an effective filtering scale.
    In Fig.~\ref{fig:P1Dspecres} the Jeans scale and the conservative estimate for the effective filtering scale is marked by dashed vertical lines.
    The true filtering scale will line somewhere between these two lines, providing a range for which data is required to resolve the $b$-value contention.

    We can now consider the number of spectra needed and their requisite S/N.
    The most recent and comprehensive \lya~forest catalog only selects 82 QSOs with a S/N of at least 5 per resolution element over the entire spectrum and a median S/N of 10 \cite{Danforth:2016}.
    The most recent calculation of the \lya~forest P1D from these data used a slightly more sophisticated masking procedure, which brought the number of QSOs down to 65 \cite{Khaire:2019}.
    Fig.~\ref{fig:P1Dsnr} shows the effect of varying S/N of the spectra on the predicted P1D. 
    Noise was added to the spectra through a Gaussian noise vector calculated from the flux variance and the desired S/N.
    The S/N of \lya~forest observations limits the lowest detectable column densities for absorbers in the spectra.
    The \lya~forest covers a wide range of column densities $10^{12} < N_{HI} < 10^{17}$ cm$^{-2}$.
    The largest catalog of \lya~absorbers covers column densities as small as $10^{12.2}$ cm$^{-2}$, and most absorbers have column densities between $10^{12.6}$ and $10^{14}$ cm$^{-2}$ \cite{Danforth:2016}.
    As lower column density absorbers are more abundant, increasing S/N will be vital to obtain a representative population of \lya~forest absorbers.
    Similarly, as the highest column density absorbers are the rarest, obtaining a large number of QSO observations will be required.
    The effect of S/N is most apparent on small-scales ($k \gtrsim 0.03$ s/km) and on large-scales the impact is less than a few percent.

    Fig.~\ref{fig:P1Dpathlength} shows variations in the predicted P1D given a different total redshift pathlength. 
    We take the ratio of the P1D from different total numbers of sightlines (calculated using bootstrapping) to that of the P1D computed from 5000 sightlines (a number found to produce a converged P1D).
    Each sightline is the length of the simulation box from which a redshift pathlength can be calculated based on the given redshift. 
    From this plot, we can determine the pathlength required to achieve a certain P1D precision, which then translates to the number of QSO observations required.
    The variation in the P1D given different pathlengths is the result of cosmic variance or the lack of a representative sample of absorbers.
    We note that this is probably an underestimation due to the finite box size (100 Mpc/h side length) of the simulations from which the sightlines are generated and that the simulation represents only one realization of the density field.
    We report the 2$\sigma$ confidence interval in Fig.~\ref{fig:P1Dpathlength} and Table~\ref{tab:pathlength} due to these uncertainties. 
    The 1$\sigma$ confidence interval from a pathlength of $\Delta z = 10$ can achieve a P1D precision of 5\% for each redshift bin given that the other requirements of S/N and $\Delta v$ are met. 
    We discuss later how using 1$\sigma$ confidence would change the results as opposed to using 2$\sigma$ confidence.

    \begin{table}[]
    \caption{Observational data requirements for P1D convergence based on the Simba simulations. Data is reported for the redshift bins shown in plots. Reported is the redshift pathlength ($\Delta z$, with 2$\sigma$ confidence), number of QSOs that corresponds to, S/N (per resolution element assuming the corresponding $\Delta v$), and spectral resolution ($\Delta v$). If requirements are met at $z= 0.7$ and $z=1.6$ for the FUV and NUV bins respectively then requirements are automatically met at the lower redshifts, assuming those QSO observations also cover the lower redshift wavelengths.}
    \label{tab:pathlength}
    \begin{center} 
        \begin{tabular}{|cccccccccc|}
        \hline
        \multicolumn{1}{|c|}{\cellcolor[HTML]{C0C0C0}} & \multicolumn{5}{c|}{\cellcolor[HTML]{C0C0C0}\textbf{FUV}} & \multicolumn{4}{c|}{\cellcolor[HTML]{C0C0C0}\textbf{NUV}} \\ \hline
        \multicolumn{1}{|c|}{$z$} & 0.03 & 0.1 & 0.2 & 0.3 & \multicolumn{1}{c|}{0.4} & 0.7 & 1 & 1.3 & 1.6 \\ \hline
        \multicolumn{10}{|l|}{\cellcolor[HTML]{EFEFEF}For 20\% P1D Precision - \textit{State of the Art}} \\ \hline
        \multicolumn{1}{|c|}{$\Delta z$ ($2\sigma$) } & \multicolumn{5}{c|}{} & 10 & - & - & 10 \\
        \multicolumn{1}{|c|}{\# of QSOs ($2\sigma$) } & \multicolumn{5}{c|}{} & - & - & - & 33 \\
        \multicolumn{1}{|c|}{S/N} & \multicolumn{5}{c|}{} & $\geq 5$ &  &  & $\geq 5$ \\
        \multicolumn{1}{|c|}{$\Delta v$} & \multicolumn{5}{c|}{\multirow{-4}{*}{Achieved with current data.}} & 15 km/s & - & - & 15 km/s \\ \hline
        \multicolumn{10}{|l|}{\cellcolor[HTML]{EFEFEF}For 10\% P1D Precision - \textit{Improved}} \\ \hline
        \multicolumn{1}{|c|}{$\Delta z$ ($2\sigma$) } & - & 25 & - & - & \multicolumn{1}{c|}{25} & 25 & - & - & 25 \\
        \multicolumn{1}{|c|}{\# of QSOs ($2\sigma$) } & - & - & - & - & \multicolumn{1}{c|}{250} & - & - & - & 84 \\
        \multicolumn{1}{|c|}{S/N} &  & $\geq 20$ &  &  & \multicolumn{1}{c|}{$\geq 15$} & $\geq 10$ &  &  & $\geq 5$ \\
        \multicolumn{1}{|c|}{$\Delta v$} & - & 10 km/s & - & - & \multicolumn{1}{c|}{10 km/s} & 7.5 km/s & - & - & 7.5 km/s \\ \hline
        \multicolumn{10}{|l|}{\cellcolor[HTML]{EFEFEF}For 5\% P1D Precision - \textit{Desired}} \\ \hline
        \multicolumn{1}{|c|}{$\Delta z$ ($2\sigma$) } & - & 75 & - & -  & \multicolumn{1}{c|}{75} & 75 & -  & - & 75 \\
        \multicolumn{1}{|c|}{\# of QSOs ($2\sigma$) } & - & - & - & - & \multicolumn{1}{c|}{750} & - & - & - & 250 \\
        \multicolumn{1}{|c|}{S/N} &  & $\geq 25$ &  &  & \multicolumn{1}{c|}{$\geq 25$} & $\geq 20$ &  &  & $\geq 10$ \\
        \multicolumn{1}{|c|}{$\Delta v$} & - & 7.5 km/s & - & - & \multicolumn{1}{c|}{7.5 km/s} & 6 km/s & - & - & 6 km/s \\ \hline
        \end{tabular}
    \end{center}
    \end{table}

    To determine the number of QSOs that must be observed given a redshift pathlength, we first split the data into two wavelength bins. 
    For $0.7<z<1.6$ the \lya~transition is shifted into the NUV.
    QSOs tend to have brighter NUV magnitudes than FUV magnitudes at these redshifts.
    For our purposes, we assume that if we observe $z=1.6$ QSOs that they can provide data down to $z=0.7$. 
    For $0<z<0.7$ the \lya~transition is in the FUV.
    Similarly to the higher redshifts, observing QSOs at $z=0.7$ should provide the required data to $z=0$.
    The number of QSOs required to meet a certain redshift pathlength is $\Delta z$ divided by the width of the redshift bin. 
    For the NUV the width of the redshift bins chosen is $\delta z = 0.3$ so the total number of QSO observations required is $\Delta z / \delta z$. 
    A similar calculation is performed for the FUV but with $\delta z = 0.1$.

\begin{figure}
        \begin{center}
        \begin{tabular}{c}
        \includegraphics[height=7cm]{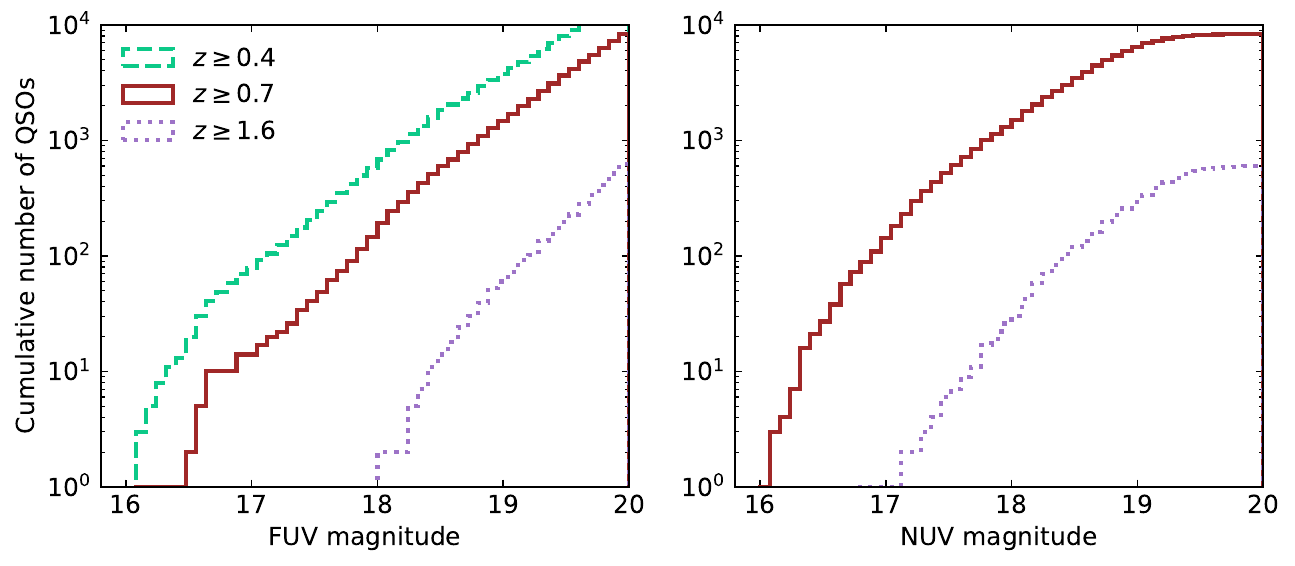}
        \end{tabular}
        \end{center}
        \caption 
        { \label{fig:FUVmag}
        The cumulative number of known QSOs reported in the Milliquas catalog\cite{milliquas} for their given FUV (left) and NUV (right) magnitudes. At $z\gtrsim0.7$ the \lya~transition starts to be observed in the NUV band. These numbers inform down to what magnitude will need to be observed in order to obtain the required number of QSO observations from Table~\ref{tab:pathlength}.} 
    \end{figure}

    We can then look at the number of QSOs from the Milliquas catalog\cite{milliquas}, a collection of known radio/X-ray associated QSO candidates  from different sources including several surveys, to determine what FUV / NUV magnitudes will be needed to collect enough data (Fig.~\ref{fig:FUVmag}).
    Basing the magnitudes off of these known QSOs will simplify data collection by aiming at existing sources rather than looking for new objects.
    This also makes calculating estimated exposure times more accurate since we know the dimmest object that might need to be observed.

    Table \ref{tab:pathlength} has a comprehensive breakdown of the observational requirements for different P1D precision levels including the redshift pathlength, total number of QSOs, the required spectral resolution, and the desired S/N.
    The results are reported for both the FUV and NUV redshift ranges and for 3 different P1D precision scenarios: 20\%, 10\%, and 5\% which we label as ``State of the Art'', ``Improved'', and ``Desired'' respectively. State of the Art corresponds to the precision we already have for the \lya~forest at $z<0.4$ and what it would take to achieve complementary data at $0.7<z<1.6$. Improved corresponds to data that will allow some constraints to be made on AGN feedback models and more precise analyses of the \lya~forest especially with the $b$-value distribution. Desired corresponds to data that can definitively distinguish between certain AGN feedback models and provides the best opportunity to conduct ground-breaking science with the \lya~forest given existing uncertainties.

    To obtain State of the Art \lya~forest data in the redshift range $0.4<z<1.6$ that complements the already existing data at $z<0.4$, P1D precision of $\sim 20$\% is required.
    This can be achieved with $\sim33$ QSO observations with S/N$\geq5$ and spectral resolution $\Delta v \leq 15$ km/s.
    This already pushes the spectral resolution below that of HST-COS and close to that of the $\Delta v = 7.5$ km/s required for transformative science regarding the thermal state of the IGM.
    Looking at the number of known QSOs from the Milliquas catalog\cite{milliquas}, these observations can be achieved with NUV magnitudes as low as 17.5. 

    To obtain Improved \lya~forest data with a P1D convergence within 10\%, which theoretically can distinguish between different AGN feedback models at $z\leq 0.7$ \cite{Tillman:2025}, more QSO observations are required.
    At $z=1.6$, about 84 QSO observations with S/N $\gtrsim 10$ and $\Delta v \leq 7.5$ km/s are required. 
    Instead, at $z=0.7$ we need $\sim 250$ QSO observations with S/N~$\geq 20$ and $\Delta v \leq 10$ km/s. 
    The necessary FUV and NUV magnitudes to obtain these numbers are $\sim 17.3$ and $\sim 17.8$ respectively.

    The Desired \lya~forest data will allow more definitive conclusions to be drawn across the entire redshift range ($0<z<1.6$), given a P1D precision of 5\%. 
    This level of precision also leaves additional room for errors due to cosmic variance or other unforeseen factors.
    This can be achieved with $\sim 250$ QSO observations with S/N~$\geq 20$ and $\Delta v = 6$ km/s at $z=1.6$.
    At $z=0.7$, about $750$ QSO observations with S/N~$\geq 25$ and $\Delta v = 7.5$ km/s are required. 
    The required FUV and NUV magnitudes for these observations given known QSO numbers are 17.6 and 18.

    As spectral resolution is the least flexible parameter given instrumentation requirements, this leaves two scenarios for \lya~forest observations with HWO.
    A $\Delta v = 10$ km/s (R$\sim$ 30,000) enables State of the Art P1D precision for $0.7<z<1.6$ and Improved precision for $0<z<0.7$.
    This will provide data at higher redshifts that complement existing data, allow for improvements on existing data at $z<0.4$, all of which will paint a clearer picture of the thermal evolution of the IGM at $z<2$. 
    Furthermore, this will allow for the use of the \lya~forest as a constraint on certain AGN feedback models.
    These results will be notable for furthering our understanding of the large-scale impacts of AGN feedback.
    However, uncertainties will remain due to numerical limitations of simulation, and the contention between simulated and observed $b$-value distributions will likely remain unresolved.
    A higher spectral resolution of $\Delta v = 7.5$ km/s (R$\sim$ 40,000) provides Improved P1D precision for $0.7<z<1.6$ and Desired precision for $0<z<0.7$.
    With these data, more AGN feedback models can be constrained using the \lya~forest.
    These data will also provide confidence in the observed $b$-value distribution that will guide the resolution of that contention between simulations and observations.
    A spectral resolution of $\Delta v = 15$ km/s (R$\sim 20000$) would severly limit the \lya~forest science we can perform using HWO but would allow the collection of data in the redshift range $0.7 < z < 1.6$ that complements existing data at $z<0.4$.

    The large number of QSO observations needed for the Improved and Desired scenarios results from the 2$\sigma$ confidence interval requirement for the pathlength estimation. 
    If we reduce the requirement to a 1$\sigma$ confidence interval, we can achieve 5\% P1D precision across all redshift ranges with a pathlength as small as $\Delta z = 10$. 
    This reduces the number of QSO observations required to 33 at $z=1.6$ and 100 at $z=0.7$ which would required a FUV/NUV magnitude of 16.6 and 17.5 respectively. 
    However, this number is less likely to account for uncertainties due to cosmic variance and DLA/LLS abundance.

\section{Discussion and Conclusion}

    For decades, the \lya~forest has been a key scientific tool for studying IGM gas, which would otherwise be invisible in the pursuit of understanding what mechanisms (both cosmological and astrophysical) set the large- and small-scale structure of our Universe.
    However, observational data at low redshifts ($z\leq1.8$) are sparse due to the limited capabilities of current space-based spectrographs. 
    At these redshifts, astrophysical processes, such as galactic feedback, are expected to play a larger role in setting the observed \lya~forest. 
    High-precision data, especially in the redshift range $0.4\leq z \leq 1.8$, a span of cosmic time for which observations are nearly nonexistent, will aid future studies in disentangling the effects of the assumed galactic feedback, the UVB, and cosmology on the predicted \lya~forest.

    In this work, we focused on the \lya~forest P1D, a diverse statistic that reveals information on the large- and small-scale structure that makes up the IGM. 
    Existing observations constrain the \lya~forest P1D within $20-25$\% at $z\leq 0.41$ \cite{Khaire:2019}.
    Studies exploring the P1D have emphasized the need for higher precision observational data, through both higher spectral resolution and S/N, to be able to clearly distinguish between different UVB and feedback models that affect the \lya~forest \cite{Khaire:2024,Tillman:2025}.
    Tight constraints on the P1D on both small and large physical scales will be important.
    On large scales, the P1D reveals information about the ionizing background and abundance of neutral hydrogen in the IGM.
    On small scales, it probes the thermal state of the IGM and the $b$-value distribution of the \lya~forest absorbers (a statistic in which current observations and simulations disagree consistently). 
    
    A 10\% convergence of the P1D can theoretically distinguish between different feedback models when estimating the uncertainty in the assumed ionizing background. 
    Achieving 5\% convergence of the P1D would aid in groundbreaking science, as it would allow controlling for additional uncertainties and be able to distinguish between more feedback models. 
    HWO, with a primary aperture of 6.5 - 8 m, could deliver this scientific return if equipped with a sufficiently high resolution spectrograph (i.e.\ $\Delta v = 7.5$ km/s or R = 40,000).
    This spectral resolution will enable resolving the lowest $b$-value absorbers which encode information on the thermal state of the IGM.
    A lower spectral resolution ($\Delta v = 10$ km/s or R = 30,000) could still constrain large-scale phenomena but potentially leave unresolved a tension between the observed and simulated $b$-value distribution that will likely remain until the next flagship space observatory.
    A spectral resolution of $\Delta v = 15$ km/s would dramatically limit the ability to do transformative science using the \lya~forest, but still allow for new observations at higher redshifts.
    
    To cover the span of cosmic time where feedback effects become most pronounced (according to our simulation), HWO's UV spectrograph should be able to observe the \lya~forest at $0 \leq z \leq 1.8$, which requires a wavelength coverage of 1215-3402 \AA. 
    To cover this wavelength range, several gratings will need to be equipped to the planned FUV spectrograph.
    To achieve the NUV wavelengths, a complementary instrument to the FUV spectrograph may even be considered.
    Additionally, the high S/N requirements for the desired observations support a case for the large aperture size which would cut exposure times essentially in half.
    
    Given existing concept models for an HWO UV spectrograph built on the LUMOS concept for LUVOIR \cite{France:2017_lumos}, we can estimate exposure time calculations for the different cases presented herein (Improved vs. Desired as labeled in Table~\ref{tab:pathlength}) for an 8-m aperture HWO, at least for wavelengths of $\leq 1900$ \AA.
    Achieving the improved \lya~forest data at $z\leq 0.4$, requires a spectral resolution of R=30,000 and a S/N $\geq 20$. 
    This S/N can be achieved for an FUV mag $\leq$ 17.3 QSO with 1 hour of exposure time.  
    Thus, the 250 QSOs required at $z=0.4$ would require a total exposure time of 250 hours.
    To achieve the desired data at $z\leq 0.4$, a spectral resolution of R=40,000 and S/N $\geq 25$ is required.  Instead, observing an FUV mag $\leq$ 17.6 QSO for 2.3 hours would achieve the required S/N. 
    Now, the total 750 QSOs required would take a total exposure time of 1725 hours.
    These total exposure times will actually be shorter since not every QSO observed will be as dim as the one used for our example calculation. 
    For example, a QSO with an FUV mag of 16.5 would only take $\sim 0.5$ hours of exposure time to achieve the desired S/N.
    
    These total exposure times are long, but the proposed work is a supporting science case, and these data will be progressively accumulated over the lifetime of the telescope and used for more than just \lya~forest science. 
    The science proposed herein, constraining AGN feedback and the thermal history of the IGM, utilizes only the \lya~lines, all additional FUV/NUV QSO absorption lines are acquired alongside the \lya~lines for free. 
    Works that utilize these lines (e.g. studies of the circumgalactic medium, LLSs, DLAs, warm-hot IGM etc.) will obtain an abundance of data for their science from these observations.
    Studies involving DLAs and LLSs could allow for an independent constraint on HI to complement 21-cm line studies \cite{Karagiannis:2022}.
    
    A flagship mission such as HWO is uniquely suited to deliver breakthrough progress for \lya~forest science.
    With the \lya~forest observations described herein, HWO can answer some of the most perplexing questions in extragalactic astrophysics regarding the role of galactic feedback in setting the physical state of the IGM. 

\appendix

\subsection*{Disclosures}

The authors declare that there are no financial interests, commercial affiliations, or other potential conflicts of interest that could have influenced the objectivity of this research or the writing of this paper.

\subsection* {Code, Data, and Materials Availability}

The Simba simulations are publicly available for download and use on their website. The FSFE package is publicly available on GitHub (\url{https://github.com/sbird/fake_spectra}).

\subsection* {Acknowledgments}
M.T.T. thanks Simeon Bird for advice on which FSFE functions to use to generate the various spectra analyzed within. 
B.B. acknowledges support from NSF grant AST-2009679.
B.B. acknowledges support from NASA grant No. 80NSSC20K0500. This research was also supported in part by the National Science Foundation under Grant No. NSF PHY-1748958.
B.B. is grateful for generous support from the David and Lucile Packard Foundation and the Alfred P. Sloan Foundation.
J.N.B. is grateful for funding support from National Science Foundation under Grant Number 2327438.
S.B. thanks the funding support from STScI through HST GO grant 14071, NSF AAG grants 2108159 and 2009409, and NASA ADAP grant 80NSSC21K0643.
The authors thank the Flatiron Institute Center for Computation Astrophysics for providing the computing resources used to conduct much of this work. The Flatiron Institute is a division of the Simons Foundation.
M.T.T. thanks the reviewers for their guidance in improving this work.


\bibliography{report}   
\bibliographystyle{spiejour}   

\listoffigures
\listoftables

\end{spacing}
\end{document}